\documentclass[prb,showpacs,
twocolumn, aps]{revtex4}

\usepackage{graphicx}
\usepackage{dcolumn}
\usepackage{bm}

\begin{document}
\newcommand*{\TTF}{TTF-TCNQ}
\newcommand*{\BS}{(TMTSF)$_2$PF$_6$}

\title{Surface characterization and surface electronic structure of
organic quasi-one-dimensional charge transfer salts}

\author{M. Sing}
\email{sing@physik.uni-augsburg.de}
\affiliation{Institut f\"ur
Physik, Experimentalphysik II, Universit\"at Augsburg, D-86135
Augsburg, Germany}

\author{U. Schwingenschl\"ogl}
\affiliation{Institut f\"ur Physik, Experimentalphysik II,
Universit\"at Augsburg, D-86135 Augsburg, Germany}

\author{R. Claessen}
\affiliation{Institut f\"ur Physik, Experimentalphysik II,
Universit\"at Augsburg, D-86135 Augsburg, Germany}

\author{M. Dressel}

\affiliation{1.\,Physikalisches Institut, Universit\"at Stuttgart,
D-70550 Stuttgart, Germany}

\author{C.~S. Jacobsen}

\affiliation{Department of Physics, Technical University of
Denmark, DK-2800 Lyngby, Denmark}

\date{\today}

\begin{abstract}
We have thoroughly characterized the surfaces of the organic
charge-transfer salts {\TTF} and {\BS} which are generally
acknowledged as prototypical examples of one-dimensional
conductors. In particular x-ray induced photo\-emission
spectroscopy turns out to be a valuable non-destructive diagnostic
tool. We show that the observation of generic one-dimensional
signatures in photo\-emission spectra of the valence band close to
the Fermi level can be strongly affected by surface effects.
Especially, great care must be exercised taking evidence for an
unusual one-dimensional many-body state exclusively from the
observation of a pseudogap.
\end{abstract}


\maketitle

\section{Introduction}
In strictly one-dimensional (1D) metals many-body theory predicts
unusual behavior of the electronic properties due to their
fundamental instability against an infinitesimal small
perturbation of the Coulomb interaction. Such systems can no
longer be described by conventional Fermi liquid (FL) theory.
Instead, the concept of a Luttinger liquid (LL) has been
introduced which is characterized by generic 1D features. These
comprise e.g. bosonic excitation modes rather than fermionic
quasi-particles, a power-law decay of the momentum integrated
spectral weight towards the Fermi energy $E_F$ or spin-charge
separation.\cite{Voit95} Most of these signatures are best seen in
the (momentum resolved) single particle excitation spectrum as
directly probed by (angle resolved) photo\-emission spectroscopy
((AR)PES). Indeed, (quasi-)1D metals were found to display marked
deviations from conventional metallic behavior using
(AR)PES.\cite{Dardel91,Hwu92,Dardel93,Coluzza93,Claessen95,Sekiyama95,Takahashi96,Gweon96,Zwick97,Zwick98,Vescoli00,Claessen02}
Basically all 1D materials studied so far show no clear Fermi
cut-off. Only recently, we obtained convincing evidence for
spin-charge separation in the charge-transfer salt {\TTF} based on
an analysis within the 1D Hubbard model.\cite{Claessen02} However,
PES is extremely surface sensitive and any deviation from
conventional metallic behavior could simply be due to the surface
being different from the bulk. Unfortunately, up to now only
little effort has been spent on the investigation of the actual
nature of the surface under study. This would be especially
important for organic materials which are particularly susceptible
to rapid photon induced decomposition in the vacuum ultraviolet
(VUV). In this paper we aim to fill this gap for {\TTF} and deal
with another charge-transfer salt, {\BS}, to exemplify the
importance of both intrinsic and extrinsic surface effects.

\section{Organic charge-transfer salts}
The organic charge-transfer salts comprise a vast variety of
molecular crystals containing almost planar organic donor and/or
acceptor molecules as essential structural building units. These
are stacked on top of each other with a possible tilt of the
molecular planes relative to the stacking direction. Several types
of stacks can occur, e.g. consisting of only one species, or with
molecules A and B alternating along one stack or being segregated
to form two types of chains as in {\TTF}. However, the interesting
electronic properties of these compounds are not of molecular
origin but arise from the interaction of adjacent molecules.
Depending on the ``side by side'' and ``face to face'' interaction
strength the crystals show predominantly one- and
two-,\cite{Torrance87} or even three-dimensional features in their
electronic behavior.\cite{Ishida01} The intermolecular interaction
involves the $\pi$ orbitals pointing perpendicular to the
molecular plane and ranges from van der Waals-type over weakly
covalent to ionic in character. An on-molecule Coulomb repulsion
energy in the range between 0.5 and 2\,eV together with the
relatively small band widths puts these systems in an intermediate
coupling regime where correlations may be
important.\cite{Torrance87} It is the quasi-tunability of the
correlation strength and the dimensionality which makes the
organic charge-transfer salts so interesting and produces this
wealth of symmetry-breaking ground states including spin- (SDW)
and charge-density waves (CDW), spin-Peierls states and even
superconductivity. In this paper we focus on two systems, {\TTF}
and {\BS} which could be classified within the above-sketched
scheme as quasi-one-dimensional mixed valency segregated stack
conductors. In {\TTF} the mixed valency is due to incomplete
charge transfer of 0.59 electrons from TTF to TCNQ while in the
so-called Bechgaard salt {\BS} it arises from the $2:1$ ratio
between the radical cation TMTSF and the counter anion
PF$_6$.\cite{Torrance87} In the following we only show data which
were recorded in the normal metallic state, i.e. above the
CDW-transition temperature of 54\,K for {\TTF} \cite{Kagoshima88}
and above the 1D-2D crossover temperature of about 110\,K for
{\BS}.\cite{Moser98}
\begin{figure}
\includegraphics[width=8cm]{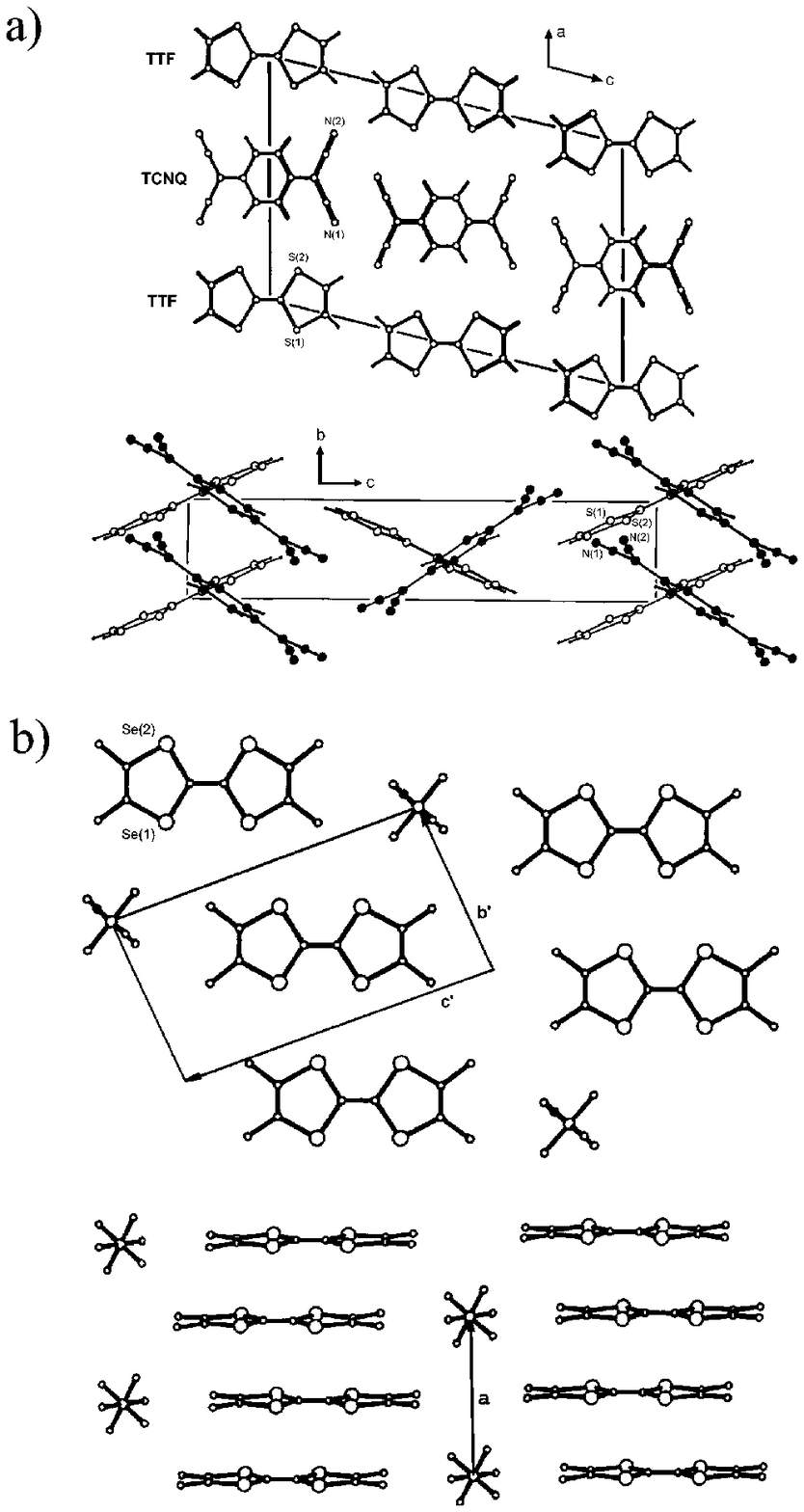}
\caption{\label{structure} (a) View of the crystal structure of
{\TTF} along the {\bf b} axis and side view of the ({\bf a}, {\bf
b}) plane (after Ref.~\onlinecite{Kistenmacher74}). (b) View of
the crystal structure of {\BS} along the {\bf a} axis and side
view of the ({\bf a}, {\bf b}) plane.} {\bf b}$^{\prime}$ and {\bf
c}$^{\prime}$ denote the projections of {\bf b } and {\bf c}
(after Ref.~\onlinecite{Thorup81}).
\end{figure}

{\TTF} (C$_{18}$H$_8$N$_4$S$_4$) crystallizes in a monoclinic
structure (Fig.~\ref{structure}), space group $P2_1/c$, with
lattice parameters $a=12.298$\,{\AA}, $b=3.819$\,{\AA},
$c=18.468$\,{\AA}, and
$\beta=104.46^{\circ}$.\cite{Kistenmacher74} The segregated TTF
and TCNQ stacks run along the crystallographic {\bf b} direction.
The molecular planes are tilted (with opposite signs) by
$24.5^{\circ}$ (TTF) and $34.0^{\circ}$ (TCNQ) with respect to
{\bf b} around {\bf a}. The two types of chains alternate along
{\bf a} while they do not along {\bf c}. Within a unit cell there
are two TTF (TCNQ) chains with opposite tilting angles of the
molecules thus leading to a herringbone-type of arrangement.

The crystal structure of {\BS} (2C$_{10}$H$_{12}$Se$_4$.PF$_6$) is
triclinic, space group $P \bar 1$, with lattice parameters
$a=7.297$\,{\AA}, $b=7.711$\,{\AA}, $c=13.522$\,{\AA}, and angles
$\alpha=83.39^\circ$, $\beta=86.27^\circ$, $\gamma=71.01^\circ$ at
300\,K.\cite{Thorup81} The easy axis, the crystallographic {\bf a}
direction, is made up by TMTSF stacks stabilized by the negatively
charged PF$_6$ counter ions in between. The molecular plane is
almost perpendicular to {\bf a}.

\section{Experimental details}
The {\TTF} and {\BS} single crystals were grown by diffusion (in
pure acetonitrile) and electro-crystallization, respectively. They
had typical dimensions of $0.8 \times 3.0 \times 0.2$\,mm$^3$ and
$1.5 \times 0.5 \times 0.1$\,mm$^3$, respectively, with their 1D
direction along the long sample axes. Their lancet-like shape
makes it only possible to perform PES measurements on the (001)
($({\bf a}, {\bf b})$) plane for both {\TTF} and {\BS}.

For both systems clean surfaces were exposed by {\it in situ}
cleavage of the crystals at a base pressure in the low
$10^{-10}$\,mbar range through knocking off a post glued on the
sample surface. PES spectra were recorded using an OMICRON
Multiprobe surface analysis system equipped with an EA~125
analyzer. For x-ray induced photo\-emission spectroscopy (XPS) the
total energy resolution was set to 0.6\,eV while for
photo\-emission in the ultraviolet (UPS) the energy resolution
amounted typically to $\approx 70$\,meV and $\approx 150$\,meV for
{\TTF} and {\BS}, respectively. The acceptance angle was $\pm
8^{\circ}$ for XPS and $\pm 1^{\circ}$ for UPS. Monochromatized
Al~K$_{\alpha}$ radiation ($h\nu = 1486.6$\,eV) and
unmonochromatized He\,{\sc I} photons (21.22\,eV) from a
conventional discharge lamp were taken as excitation sources.
Calibration of the binding energy scale was achieved by measuring
the Fermi edge of a freshly sputtered Au foil at low temperatures.
All XPS spectra were recorded at room temperature whereas the UPS
spectra on {\TTF} and {\BS} were taken at 60\,K and 150\,K,
respectively.

\section{TTF-TCNQ}
\subsection{Ideal and actual sample surface in direct space}
First of all it is important to note that the natural cleavage
plane of {\TTF} is parallel to the (001) lattice plane. If we
regard the extended molecules for a moment as represented by point
charges it is immediately seen that this (001) {\it lattice plane}
essentially bears no net surface charge since it contains as many
TTF- as TCNQ- molecules (Fig.~\ref{structure}). Thus there is no
charge imbalance and the surface created by exposing this plane
should essentially be stable. Taking into account more
realistically the planar shape and the bulk arrangement of the TTF-
and TCNQ-molecules the same holds for the (001) {\it layer}.
However, due to the broken translational symmetry the Madelung
potential at the surface will differ from that in the bulk.
\begin{figure}
\includegraphics[width=8cm]{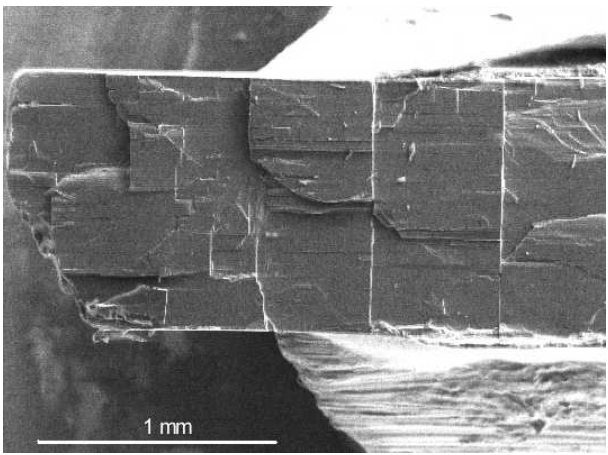}
\caption{\label{TTF-REM}SEM image
of a typical {\TTF} surface after cleavage.}
\end{figure}
Thus it is conceivable that there will occur some electronic charge
redistribution probably concomitant with a structural surface
relaxation. Since the intramolecular covalent bonds are quite
strong and hence the molecules themselves rigid and since, in
addition, there are no dangling bonds perpendicular to the surface
such a structural relaxation most likely will happen by changes in
the tilting angles with respect to the bulk. One could speculate
that structural changes at the surface will take place such that a
better screening of the Madelung potential is achieved, i.e. by a
stronger hybridization of the $\pi$ orbitals perpendicular to the
plane of the molecules. These ideas will be discussed in more
detail below.

Figure~\ref{TTF-REM} shows a typical scanning electron microscopy
(SEM) image of a cleaved {\TTF} crystal. One clearly sees the good
quality of the exposed surface with large flat terraces. Thus, the
actual sample surface indeed can be viewed as representing the
(001) lattice plane.

\subsection{Surface characterization by XPS}
\begin{figure}
\includegraphics[width=8cm]{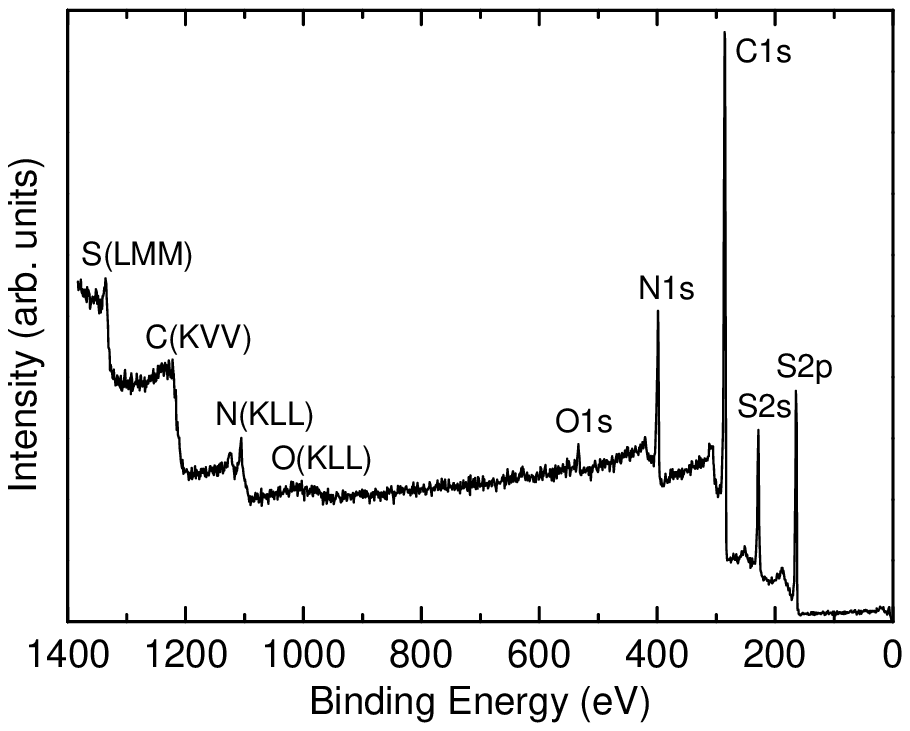}
\caption{\label{XPSTTFoverview}XPS overview spectrum of a {\TTF}
surface as exposed by {\it in situ} cleavage of a single
crystalline sample.}
\end{figure}

The surface composition of the {\TTF} crystals was investigated by
means of XPS. Figure~\ref{XPSTTFoverview} shows an overview
spectrum of a {\TTF} surface. Each spectral feature in the
spectrum can be identified and classified according to its
physical origin, i.\,e. as stemming from core levels or Auger
processes. In addition, one can find satellite structures to each
intense core level at multiples of about 22\,eV away from the main
line. These are related to inelastic losses suffered by the
photoelectrons due to plasmon excitations of all the valence
electrons. Except for a slight O contamination (see below) we find
only signatures of the constituent elements of {\TTF}.

For a quantitative analysis it is important to assure that the
individual core lines and their plasmon satellites are well
separated from each other so that there is no contribution of
other origin except for a structureless background due to
secondary electrons. In order to determine the spectral weight of
a certain core excitation a Shirley background was subtracted
before integration. The areas thus obtained were weighted by the
inelastic mean free path of the photoelectrons (which is a
function of kinetic energy), the transmission function of the
analyser (also a function of kinetic energy), and the
photoexcitation cross sections. Using
experimentally\cite{Wagner81} and theoretically\cite{Yeh85}
derived cross sections basically yields identical results. In
Tab.~\ref{TTFstoich} we summarize values obtained for a typical
sample in normal-emission (NE) geometry employing the
experimentally determined cross sections. Note that the error
amounts to about 20\,\%, mainly due to the uncertainty of the
tabulated cross sections used. Nevertheless, the agreement of the
surface composition as determined by XPS and the nominal
composition given by the bulk stoichiometry of the material is
striking. Especially there is no excess of carbon detectable and
only a weak contamination with oxygen is observed.

\begin{table}
\begin{ruledtabular}
\begin{tabular}{lccccc}
element [core line] & O~[1$s$] & C~[1$s$] & N~[1$s$] & S~[2$s$]\\
\hline nominal composition & 0 & 36 & 8 & 8\\ from XPS & 0.4 &
34.2 & 7.8 & 8
\end{tabular}
\end{ruledtabular}
\caption{\label{TTFstoich}Surface composition of {\TTF} as derived
from a quantitative analysis of the XPS core level spectra.
Experimental compositions are given with respect to sulphur. The
values have to be read as numbers of atoms per unit cell.}
\end{table}

\begin{figure}
\center{\includegraphics[angle=-90,width=8cm]{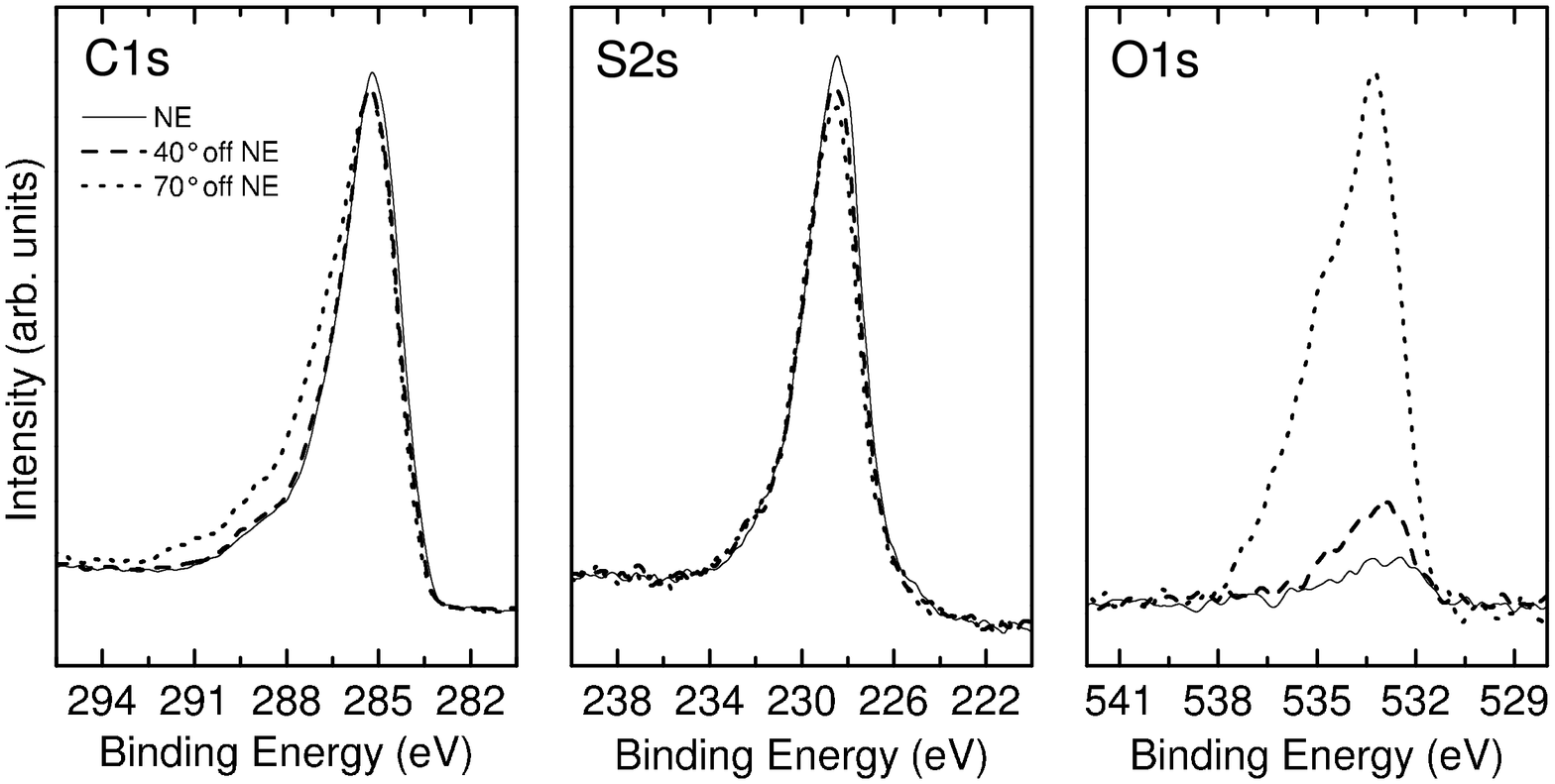}}
\caption{\label{TTF_C1s_S2s_O1s} XPS spectra of the C~1$s$,
S~2$s$, and O~1$s$ core levels of {\TTF} as a function of emission
angle.}
\end{figure}

Additional information beyond a qualitative and quantitative
elemental analysis as discussed so far may be extracted from the
line shapes and the fine structure of a certain core level. We
first turn to the C~1$s$ and S~2$s$ lines since they can be
discussed on equal basis. Their XPS-spectra recorded in NE
geometry are displayed in Fig.~\ref{TTF_C1s_S2s_O1s}. The C~1$s$
and S~2$s$ lines both consist of one single peak with an
asymmetrically decaying tail at the higher binding energy side.
From the peak maxima we derive a binding energy of 285.2\,eV and
228.4\,eV for the C~1$s$ and S~2$s$ level, respectively. The
corresponding line widths (FWHM) amount to about 2.3\,eV and
2.7\,eV.
Spectra of the above mentioned core levels are rarely discussed in
the literature for {\TTF}.\cite{Butler74,Fraxedas98} As for the
C~1$s$ level the reason is obvious. Carbon is not specific for
either the TTF or the TCNQ molecule, and there are many (9)
inequivalent sites in the crystal structure. Since at each of
these the chemical environment is different the corresponding
C~1$s$ signals are shifted in energy relative to each other.
However, due to the finite experimental resolution and the
lifetime broadening of the photo\-emission final states they
overlap to one single relatively broad line as seen in
Fig.~\ref{TTF_C1s_S2s_O1s}.

A closer look at the S~2$s$ line seems to be more promising.
Sulphur is specific for the TTF molecule and there are only two
crystallographically different sites in a ratio $1:1$.
Nonetheless, these cannot be resolved (s.
Fig.~\ref{TTF_C1s_S2s_O1s}). However, provided that there exist no
further lines, e.g. due to a surface species with different
binding energy and relative intensity, the superposition of two
symmetric line shapes contributing with equal strength is always
symmetric. Thus, from the S~2$s$ spectrum we conclude that the
asymmetric tail indeed is inherent to each single component.

\begin{figure}
\center{\includegraphics[angle=-90,width=8cm]{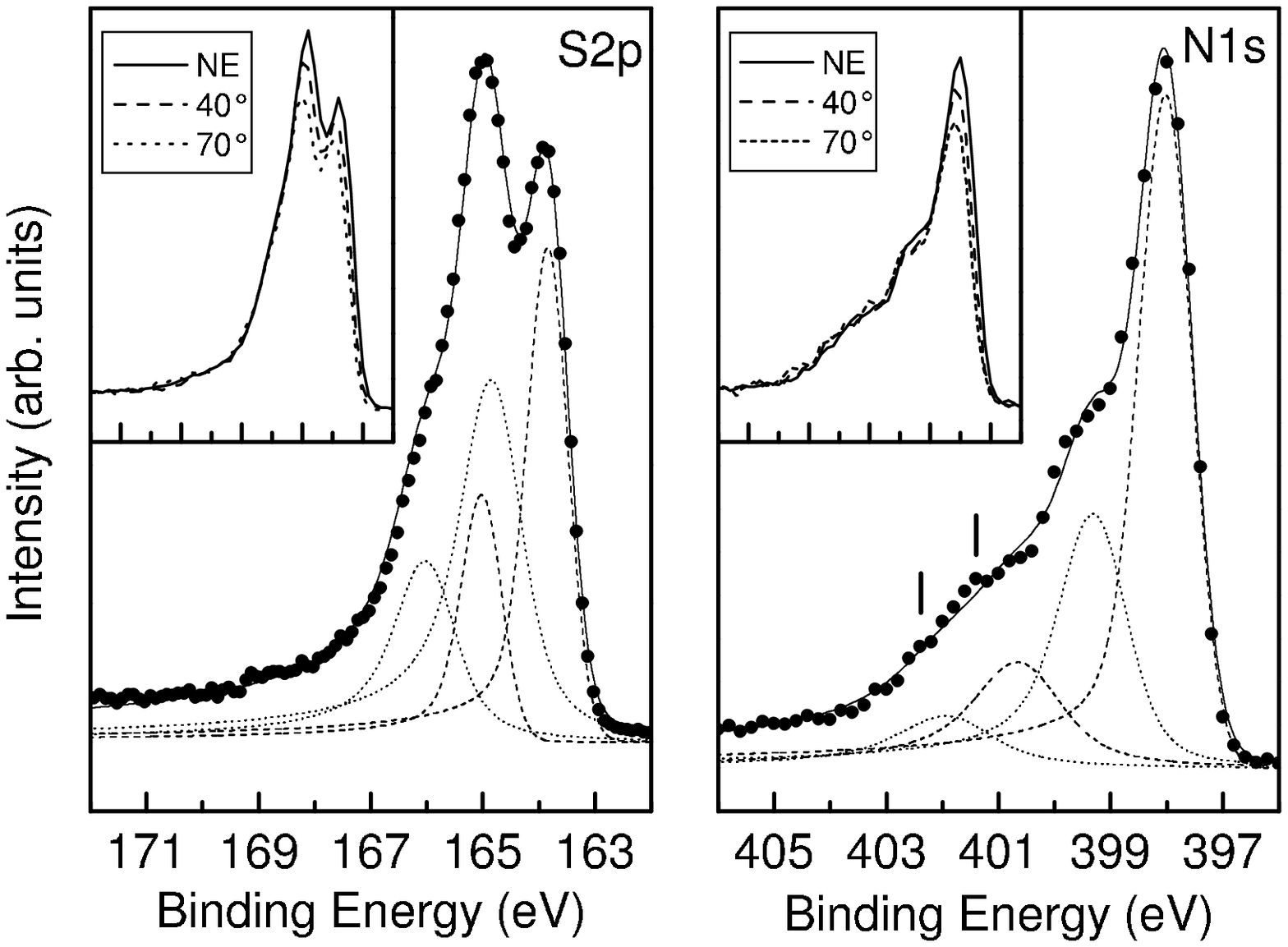}}
\caption{\label{TTF_S2p_N1s_fit}XPS spectra of the S~2$p$ and
N~1$s$ core levels of {\TTF} (dots). The lines represent
decompositions into underlying components obtained by a
least-squares fit. For details see the text. Insets: XPS spectra
of the S~2$p$ and N~1$s$ core levels as a function of emission
angle (NE, 40$^{\circ}$ off NE, 70$^{\circ}$ off NE). Note that
the binding energy scale is the same as in the parent plot.}
\end{figure}

That the situation actually is more subtle can be seen from the
S~2$p$ line shown in Fig.~\ref{TTF_S2p_N1s_fit}. In contrast to
the S~2$s$ line, the S~2$p$ signal is split into two maxima at
about 163.8\,eV and 164.8\,eV. The former maximum is about 15\%
higher in intensity than the latter. An additional shoulder is
situated at about 165.9\,eV. Again an asymmetrically decaying tail
is seen at higher binding energies. It is obvious that the two
maxima about 1\,eV apart cannot be identified with the spin-orbit
split 2$p$ doublet. They exhibit not only a quantitatively wrong
intensity ratio (expected to be $2:1$ between lower and higher
binding energy peak), but it is even reversed with the lower
binding energy peak being significantly weaker. Furthermore, the
splitting of the order of 1\,eV seems far too high to be explained
by a chemical shift of the binding energies due to the two
inequivalent S sites. The bonding lengths of the S(1) and S(2)
atoms (s. Fig.~\ref{structure}) are almost
equal.\cite{Kistenmacher74} Also the intermolecular environment of
the S(1) and S(2) atoms is topologically similar and in particular
exhibits similar distances of the S sites to the neighboring TCNQ
molecules. One has to conclude that there exist two significantly
different S signals with possibly different relative strength
which questions the above reasoning regarding the S~2$s$ line.
Thus, in order to clarify the situation it is necessary to perform
a line shape analysis of the S~2$p$ line. In accordance with our
conclusion above we used two doublets with the spin-orbit
splitting (intensity ratio $2:1$) fixed at 1.18\,eV.
\cite{Moulder92} If the observed asymmetric tail is intrinsic for
each core level and not only caused by a superposition of
different lines it would be readily explained by collective
screening of the conduction electrons as is well known for
metals.\cite{Doniach70} Hence, to each component in our analysis
we assigned the so-called Doniach-\v{S}unji\'{c} line shape
describing the metallic screening. Besides the energy position and
width a parameter $\alpha$ enters its definition which determines
the asymmetry. A Lorentzian is recovered for $\alpha = 0$. We used
one single $\alpha$ for all components. In the fitting procedure
included was also a convolution by a Gaussian of variable width to
account for the experimental resolution ($\approx 0.6$\,eV) and
contributions to the line width which do not stem from purely
exponential decay e.g. due to the coupling to phonons. Allowing
for larger Gaussian widths than justified by the experimental
resolution alters the line shapes of each component in that the
onset at lower binding energies gets steeper, i.e. more
Gaussian-like in character. The overall width and the peak
asymmetry as well as all other fit parameters remain essentially
unchanged. The results are displayed in
Fig.~\ref{TTF_S2p_N1s_fit}. The experimental spectrum is
reproduced very well. The asymmetry parameter $\alpha$ comes out
to be 0.11, in reasonable agreement with values for simple
metals.\cite{Huefner96} From this analysis we infer the intrinsic
character of the asymmetric tail due to the coupling of the
photohole to the conduction electrons and confirm that essentially
two S signals are observed which, as stated above, cannot be
reconciled by the chemical shift of the binding energies of the
two inequivalent S sites. The most interesting quantity to be
explained is the intensity ratio of the lower to the higher
binding energy contribution, which from the fit is found to be
$0.44:0.56$.

Before further elucidating the origin of the two components of the
S~2$p$ line and their intensity ratio we first turn to the line
most intensively discussed in the literature,
\cite{Grobman74,Epstein75,Swingle75,Metzger75,Epstein76,Grobman76,Ritsko78,Fraxedas98,Rojas01}
the N~1$s$ core level excitation. As is obvious from
Fig.~\ref{TTF_S2p_N1s_fit} it consists of at least three
contributions, a distinct maximum at about 398.0\,eV and two
shoulders at higher binding energies of about 399.5\,eV and
401.4\,eV. Note that for similar arguments as in the case of
sulphur these energy differences are too large to be accounted for
by possible chemical shifts of the two inequivalent N sites.
Looking closer at the shoulder at highest binding energy one can
actually distinguish additional fine structure which may be
connected to two underlying components (marked by ticks in
Fig.~\ref{TTF_S2p_N1s_fit}). Since they appear to be equally
spaced and to display a similar intensity ratio as the two
structures at lower binding energy we identify them simply as
accompanying satellite features of two different components. This
assignment is in line with the N~1$s$ spectrum of pure TCNQ
crystals, which consists of one main line and a satellite
structure well separated by about 2.6\,eV.\cite{Lindquist88} This
satellite was attributed to an intramolecular shake-up process
between the highest occupied molecular orbital of the neutral and
the lowest unoccupied molecular orbital of the ionized TCNQ
molecule. Due to the only weak covalent bonding similar local
excitations will persist in {\TTF}.

We fitted this model to our data where again we simulated the line
shapes according to Doniach and \v{S}unji\'{c}. This applies also
to the satellites although their actual spectral form is
microscopically of other origin. Without any additional information
the two main lines and one of the satellite peaks (the more
pronounced one at lower binding energy) have to be varied
independently while the second satellite is coupled to its main
line in the same way, i.e. with respect to energy position, width,
and weight, as the first one. In addition, we employed for
simplicity only one single asymmetry parameter. The results of the
fit are displayed in Fig.~\ref{TTF_S2p_N1s_fit}. The important
quantities we can extract are the main line--satellite splitting of
about 2.6\,eV, the asymmetry parameter $\alpha=0.11$ and the
intensity ratio of about $0.65:0.35$ between lower and higher
binding energy contributions.

We note that the 2.6\,eV main line--satellite splitting perfectly
agrees with the experimental value for pure TCNQ and thus confirms
our fit model. It is now interesting to correlate the intensity
ratio for lower and higher binding energy contributions with that
obtained for S~2$p$. Intriguingly, the ratios have within the
accuracy of this evaluation just reciprocal values. The idea
suggests itself that this may have something to do with the
electron transfer from TTF to TCNQ. This leaves the TTF and TCNQ
molecules in a mixed valent state of 0.59$^+$ and 0.59$^-$,
respectively. If the charge fluctuations between TTF$^0$ and
TTF$^+$ on the one hand and TCNQ$^0$ and TCNQ$^-$ on the other
take place on a slower time scale than the photo\-emission process
itself one would observe two peaks corresponding to the two
chemical states of TTF and TCNQ, respectively. Moreover, due to
less effective screening of the core potential the TTF$^+$ state
should show up in the S~2$p$ spectrum at higher binding energy
compared to the neutral chemical state. The reverse is true for
TCNQ$^-$ and the N~1$s$ line. In both cases the charged state
should have a larger spectral weight with a ratio $0.59:0.41$.
Indeed, this scenario matches qualitatively our data and is even
in fair quantitative agreement with our line shape analysis.

We only briefly mention here the controversial debate regarding the
correct interpretation of the N~1$s$ spectral features in the 70's.
Partly, it was caused by the minor quality of the data which showed
quite large intensity variations depending on the method of sample
preparation.\cite{Grobman74,Swingle75,Ritsko78} In particular, none
of these measurements were done on cleaved single crystals as in
this work. Thus, a reliable quantitative analysis was highly
impeded although the idea of two chemical states of N to be seen in
the spectra was used early in order to determine the amount of
charge transfer.\cite{Grobman74} Moreover, much of the persuasive
power of our above argumentation is owed to the correlation of the
results of our analysis for the N~1$s$ and S~2$p$ spectra which
previous work failed to attempt.\cite{Ikemoto77,Tokumoto82}
Instead, it was argued from calculations of the Madelung potentials
that Coulomb energy differences may account for the observed
relative energy shifts.\cite{Epstein75,Metzger75,Epstein76}
However, it was shown that as no polarization effects in the solid
state were taken into account such calculations were of little
use.\cite{Grobman76}

Not least because of some reports on evidence for strong angle
dependent intensity variations in XPS spectra of {\TTF}, especially
regarding the N~1$s$ level, \cite{Swingle75,Ritsko78} some space is
given to that issue here. Since 95\,\% of the detected
photoelectrons at a certain kinetic energy which were not scattered
inelastically stem from a layer of thickness $\sim 3
\lambda {\rm cos} \theta$, where $\lambda$ is the inelastic mean free
path at that energy, the information depth of XPS can be varied on
a scale of about $\sim 30$\,{\AA} by changing the detection angle
with respect to the surface normal. The results of our measurements
at 0$^{\circ}$ (NE), 40$^{\circ}$ and 70$^{\circ}$ are displayed as
solid, dashed and dotted lines in Fig.~\ref{TTF_C1s_S2s_O1s} and
the insets of Fig.~\ref{TTF_S2p_N1s_fit}. The spectra are
normalized to the background intensity at low binding energies.
Note that the background might be angle dependent as well. Hence,
only pronounced intensity variations as a function of emission
angle should be taken seriously. In view of this caveat the S~2$s$,
S~2$p$, and N~1$s$ lines are not conspicuous. The slightly
decreasing peak heights with increasing emission angles are most
probably just a matter of the normalization being systematically
wrong. On the contrary, the C~1$s$ line displays at the biggest
emission angle some additional spectral weight at higher binding
energies. This is likely due to a slight surface contamination.
Remarkable, however, is the angular dependence of the O~1$s$ line.
While only a weak signal is seen at NE and 40$^{\circ}$ off NE a
strongly enhanced peak emerges at 70$^{\circ}$ off NE. This
behavior provides striking evidence that the O must be accumulated
on the topmost surface layer. It originates probably from the
residual gas molecules in the vacuum chamber. The observed O
contamination takes place on a very short time scale and has
saturated within minutes. However, since the amount is small it
does not severely affect the UPS measurements discussed below.

To make the comprehensive discussion of the XPS spectra conclusive
with respect to our aim, i.e. to relate surface and electronic
structure, we summarize the results of this paragraph as follows:
XPS is a valuable diagnostic tool for the characterization of the
surfaces of the organic charge-transfer salt {\TTF}. Both the
elemental and line shape analysis point to the fact that we are
dealing with perfectly reproducible, well-defined, and hence
intrinsic surfaces of metallic character. Determination of the
charge transfer per molecule at the surface provides no hint for a
significant deviation with respect to the bulk. However, so far
nothing is anticipated regarding the question, if the surface
electronic properties are really representative for the bulk
material.

\subsection{Crystalline surface order and ARPES}
From the above paragraph we know that the chemical composition of
the surfaces under investigation is stoichiometric. In addition
the line shape analysis indicates a metallic surface character.
This may hint at long-range crystalline order. To validate this
conjecture the method of choice is diffraction with low energy
electrons (LEED). This probes the surface atomic order on a
lateral scale given by the coherence length which amounts to
typically 100\,{\AA}. Our attempts to obtain a LEED pattern
failed, however. In the light of various scanning tunneling
microscopy (STM) studies on both TTF-TCNQ films\cite{Ara95} and
as-grown crystal surfaces\cite{Sleator88,Nishiguchi98} we ascribe
this lack of observation to the destruction of the ordered surface
by the electron beam itself.
\begin{figure}
\includegraphics[width=8cm]{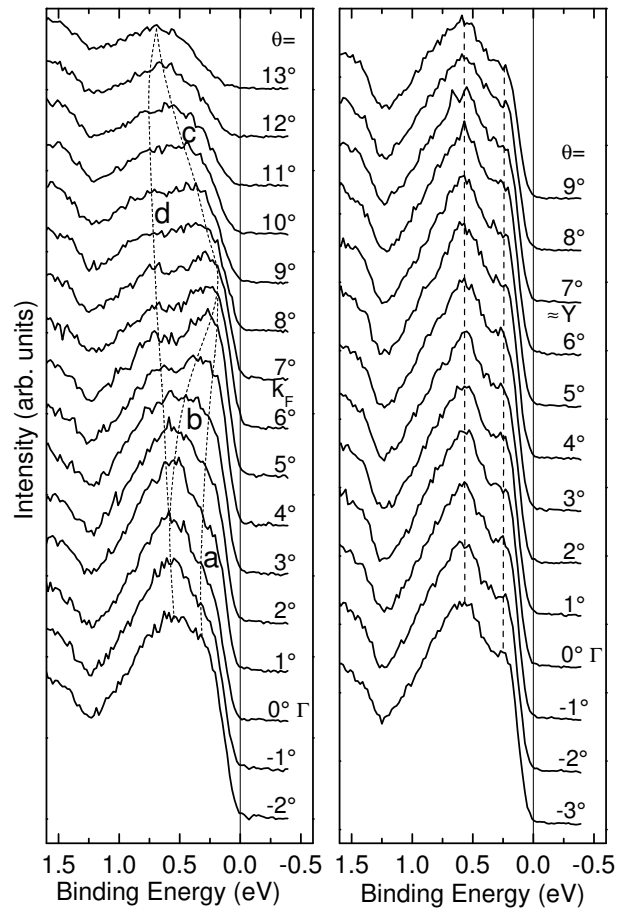}
\caption{\label{ARPES}ARPES spectra of {\TTF} along the 1D axis
(left panel) and perpendicular to it (right panel). The dashed
lines are intended as a guide to the eye. For details see the
text.}
\end{figure}
That the TTF-TCNQ surfaces indeed are long-range ordered can be
inferred from the ARPES measurements depicted in Fig.~\ref{ARPES}.
The left-hand panel shows angle-resolved measurements along the
${\Gamma}$Z-direction, i.e. along the one-dimensional {\bf b}
axis, whereas the series of the right-hand panel was recorded
perpendicular to it (including the ${\Gamma}$-point). At the
${\Gamma}$-point two peaks are observed at about 0.19\,eV
(marked~a) and 0.54\,eV (marked~b), respectively. Looking first at
the left-hand panel of Fig.~\ref{ARPES} one can follow the
dispersion of these two features (the dashed lines are intended as
a guide to the eye) both approaching the Fermi energy at an angle
around 7$^{\circ}$. Two other features can be identified. Feature
c disperses away from the Fermi energy starting at an angle of
about 7$^{\circ}$ while feature d seems to be split off feature b
at the ${\Gamma}$-point and moves to higher binding energies with
increasing angles (see dashed lines).

Switching to the right-hand panel, containing data measured
perpendicular to {\bf b}, a completely different behavior is
observed. If one follows again peaks a and b as a function of
emission angle starting with the spectrum at the ${\Gamma}$-point
essentially no dispersion is observed. The pronounced dispersions
along the 1D direction clearly indicate long-range surface order.
These together with the lack of any dispersion perpendicular to
the 1D axis on the other hand truly reflect the electronic 1D
character of the TTF-TCNQ surfaces. We refrain here from a
detailed discussion of the observed dispersions along the {\bf b}
axis. We just note that we could demonstrate previously that they
can be reconciled within the one-dimensional Hubbard model. Thus
the data bear evidence for spin-charge separation where feature a
represents the spinon and feature b the holon branch of the
excitation spectrum.\cite{Claessen02}

Rather another issue from our previous work we like to stress
here. A comparison of the ARPES derived bands with the results of
band calculations based on density functional theory (DFT) showed
experimental band widths being larger by about a factor of 2.
Otherwise good agreement of the DFT results with bulk properties,
e.g. regarding the Fermi vectors as reflected in the periodicity
of the CDW, were taken as evidence for a renormalization of the
hopping integral $t$ and hence the band width at the surface. A
possible explanation for the mechanism behind the renormalization
might be the following. In the bulk the relatively rigid TTF and
TCNQ molecules are tilted in opposite directions around the {\bf
a} direction by 24.5$^{\circ}$ and 34.0$^{\circ}$, respectively.
At the surface the Madelung potential is different from the bulk
and hence the balance between Coulomb and hybridization
interaction may readjust. This most likely involves different
tilting angles for the TTF and TCNQ molecules. Indeed it was shown
that {\TTF} films sublimed onto mica as a substrate exhibit two
kinds of phases.\cite{Ara95} One of them was identified with that
as known also from STM measurements on crystal surfaces. The other
was interpreted with a rearrangement of at least the TCNQ
molecules such that they are oriented steeper with respect to the
surface. It was argued that both arrangements deviate only
slightly in energy from each other. Since it is difficult to
determine the actual tilting angles from the STM images if
possible at all, it might be well the case that the phase only
seen on evaporated thin films is the stabilized bulk phase while
the other represents the reconstructed surface of single crystals.
We conclude that regardless of the actual reconstruction intrinsic
surface effects are important in {\TTF} and reflected in the
electronic structure of the surface.

\subsection{VUV-radiation induced surface damage}
Radiation induced surface damage both in the VUV and x-ray regime
is a well-known but rarely talked about phenomenon in the context
of PES. This is due to the fact that any time-dependent spectral
changes regardless of their origin are usually unwanted since in
most cases they signal some kind of surface degradation and hence
hinder the observation of intrinsic surface properties. Up to now
only in cases of technological interest such as in the field of
polymers there exist a number of systematic studies related to
this problem.\cite{Briggs90} Nevertheless, for several other even
inorganic materials such effects have been reported, in particular
at low temperatures.\cite{Purdie99} In any case it is important to
be aware of this issue, especially using synchrotron radiation
where the high photon flux may reduce the time scale on which
surface damage occurs down to seconds. In the following we will
address some of these aspects for {\TTF}.

\begin{figure}
\center{\includegraphics[angle=-90,width=8cm]{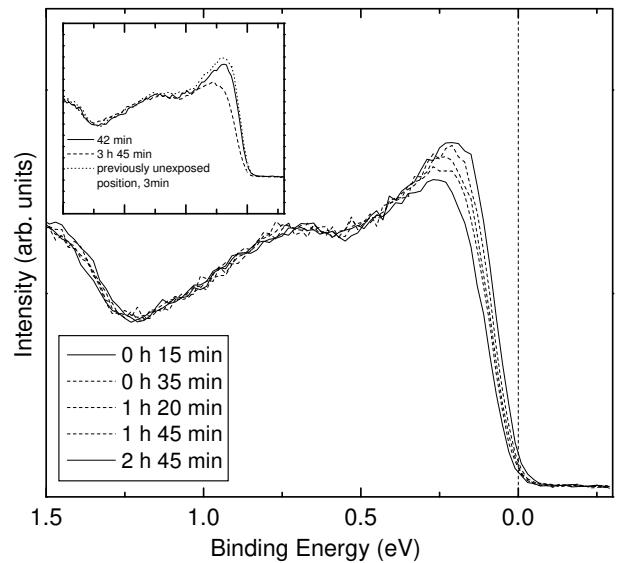}}
\caption{\label{radiation} Effects of VUV radiation on
ARPES-spectra at k$_F$. The binding energy scale of the inset is
the same as in the parent plot. For details see the text.}
\end{figure}

Figure~\ref{radiation} displays PES spectra taken at the Fermi
vector k$_{F}$ and 60\,K using the He\,{\sc I} radiation
(21.22\,eV) of a conventional unmonochromatized hollow-cathode
discharge lamp. For each curve the total VUV exposure until the
spectrum was recorded is indicated. The observed spectral changes
are twofold. First, the intensity of the structure at $E_F$
significantly decreases upon radiation exposure on a time-scale of
about 2--3~hrs. Secondly, also the energy position of this
structure changes. It shifts by about 50\,meV to higher binding
energies. That these time-dependent changes are really
radiation-induced is shown in the inset of
Fig.~\ref{radiation}. There it is demonstrated that the spectrum
taken on a freshly cleaved crystal after 42\,min VUV exposure is
fully recovered even after about 4~hrs, if one measures a
previously unexposed sample spot. Measurements using synchrotron
radiation (not shown) reveal that these degradation effects are
predominantly dependent on the photon energy (and not so much on
the intensity). Using slightly higher photon energies (25\,eV) the
tolerable VUV exposure time does not scale with the photon flux
compared to the measurements in the lab while 35\,eV photons damage
the surface within minutes. From this we conclude that there exists
a threshold or a resonance energy in the VUV regarding beam damage.
The observation that the electron beam of a LEED optics with
typical energies above 25\,eV destroys the surface almost instantly
points to the scattering of the photoexcited electrons rather than
to the photoabsorption process itself as the genuine cause for the
observed damages. Note that the observed energy shift of the
spectral feature at the Fermi energy is intrinsic and not caused by
surface charging. From conjugated $\pi$-electron systems such
irradiation effects are well-known and were attributed to the
generation of structural and chemical defects, i.e. bond breaking
and/or cross-linking.\cite{Koch01} These defects hinder the
formation of delocalized molecular $\pi$-orbitals and thus affect
first the corresponding states close to $E_F$.

We make a short remark regarding a more sophisticated explanation
of the observed phenomena. We start from the microscopic physical
picture of the undestroyed surface in terms of the 1D Hubbard
model.\cite{Claessen02} Since the 1D Hubbard model and the LL
picture are asymptotically equivalent within certain limits
\cite{Voit95} it is tempting to discuss irradiation damage under
the notion of the so-called bounded LL.\cite{Voit00,Meden00} There
the effect of finite chain length onto the spectral properties of
a LL is treated. It is conceivable that the irradiation induced
defects are local in nature and just have the effect of cutting
off the 1D chains. Introducing more and more defects means a
continuous decrease of the mean chain length. Hence the spectral
changes upon VUV irradiation would reflect the crossover to a
bounded LL. Clearly, this issue demands further exploration.

\section{{\BS}}
\subsection{Ideal and actual sample surface in direct space}
In the case of {\TTF} we have seen that following simple
considerations regarding the net charge of the exposed cleavage
surface one already gets a clue of how the surface eventually will
behave in terms of reconstruction. Our heuristically deduced
findings were confirmed by STM imaging and ARPES measurements. The
main point was to realize that the natural cleavage plane exposes
non-polar surfaces. The situation is different for {\BS}.
\begin{figure}
\includegraphics[width=8cm]{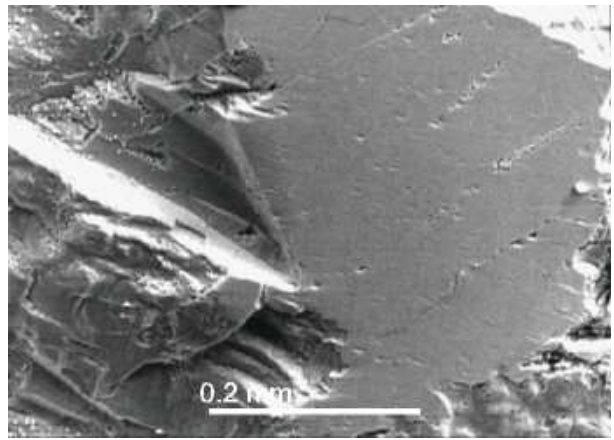}
\caption{\label{BS_REM}SEM image of
a typical {\BS} surface after cleavage.}
\end{figure}
Here the natural surface of as grown crystals is parallel to the
(001) plane. The topmost surface layer contains either only TMTSF
molecules or PF$_6$ counter ions. Thus it clearly bears a positive
or negative net surface charge (cf. Fig.~\ref{structure}) which is
energetically highly unfavourable and makes the surface especially
susceptible to electronic or atomic reconstruction. The former
possibly would lead to a modified charge transfer at the surface
changing the electronic properties severely with respect to the
bulk. On the contrary, the latter probably would induce quite a
high defect density, if there is no easy and unique way to
rearrange the surface molecules such that a distinct energy
minimum is achieved. Moreover, the polar character of the (001)
lattice plane means that in a sense there is no well-defined
natural cleavage plane. Instead of cleaving the crystal one will
rather rip it off between the (001) lattice planes. It is
conceivable that the obtained surfaces will at least be rough and
resemble more a fractured surface than being shiny and flat.
Actually, this is what we see in an SEM micrograph of an {\it in
situ} cleaved crystal (s. Fig.~\ref{BS_REM}). However, this does
not exclude the possibility of finding areas which are with or
without reconstruction undisturbed and well ordered on an atomic
scale. Indeed, STM images were reported showing a regular
arrangement of molecules.\cite{Fainchtein90} However, non-local
probes will average over macroscopic length scales and hence may
yield another picture.

\subsection{Surface characterization by XPS}
\begin{figure}
\includegraphics[angle=-90,width=8cm]{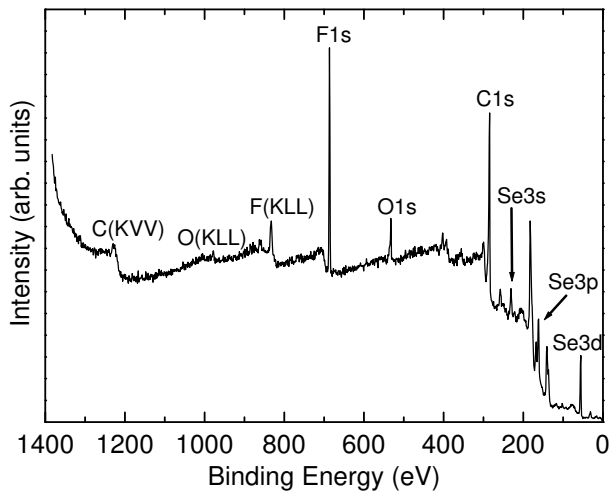}
\caption{\label{XPSBSoverview}XPS overview spectrum of a {\BS}
surface as exposed by {\it in situ} cleavage of a single
crystalline sample.}
\end{figure}

It was just shown that the surfaces of our cleaved {\BS} crystals
are rather rough compared to the ones of {\TTF} and thus might
hinder the observation of dispersing electronic states by means of
ARPES. However, the chemical composition should be unaffected by
the surface morphology. Again we used XPS for the analysis of the
surface stoichiometry. An XPS overview spectrum is displayed in
Fig.~\ref{XPSBSoverview}. The most important lines are labeled
according to their physical origin. Note that the P~2$s$ and
P~2$p$ core levels interfere with various Se-Auger features and
thus cannot be clearly discriminated. Otherwise, every line in the
spectrum can be identified. Except for C and O (see below) only
elements which are constituents of {\BS} are found. In addition to
the main lines plasmon loss features are found corresponding to a
plasmon excitation energy of about 22\,eV similar to the value
seen in {\TTF}.
Due to the overlap of various lines only a limited number of core
levels was suited for the determination of the surface composition
using the same evaluation method as above for {\TTF}. The results
for the cleavage surface which displayed the weakest O signal are
summarized in Tab.~\ref{BSstoich}.\cite{ann1} We give here the
mean of the values which one gets using the cross sections of both
Ref.~\onlinecite{Wagner81} and \onlinecite{Yeh85}, respectively.
Compared to {\TTF} the discrepancy between the nominal values and
those derived from XPS is striking. It amounts to almost 80\,\%
excess of carbon and about 40\,\% deficiency of Se.
\begin{table}
\begin{ruledtabular}
\begin{tabular}{lccccc}
element [core line] & O~[1$s$] & C~[1$s$] & Se~[3$d$] & F~[1$s$]\\
\hline nominal composition & 0 & 20 & 8 & 6\\ from XPS & 3.3 &
35.5 & 4.5 & 6
\end{tabular}
\end{ruledtabular}
\caption{\label{BSstoich}Surface composition of {\BS} as derived
from a quantitative analysis of the XPS core level spectra.
Experimental compositions are given with respect to fluorine. The
values have to be read as numbers of atoms per unit cell.}
\end{table}
Moreover, a non-negligible amount of oxygen is observed. We note
that all surfaces were freshly prepared. It should be added that
the measured compositions of the investigated surfaces scattered
unsystematically with relative deviations from the averaged values
of Tab.~\ref{BSstoich} by up to 50\,\% in contrast to the case of
{\TTF}. One thus could be led to suspect that this just reflects
the bad quality of our crystals in general. However, measurements
of the dc and microwave resistivity as well as electron spin
resonance (ESR) data on our samples neatly show the SDW transition
at 12\,K.\cite{Dumm00} The pronounced deviation of the surface
composition from the nominal one may be explained by severe
reconstructions of at least parts of the surface due to its polar
character. Just as well it could be related to processes taking
place already during crystal growth, e.g. to the substitution of
Se by the chemically equivalent O from the solvent or to
microscopic cracks or precipitations (cf.~Fig.~\ref{BS_REM}) which
are chemically modified. In any case, already from the XPS
elemental analysis we must conclude that the surfaces of {\BS} as
exposed by {\it in situ} cleavage of well-characterized single
crystals are not only {\it not} representative for the bulk
material, they even are not intrinsic surfaces.

\begin{figure}[h!!]
\includegraphics[angle=-90,width=8cm]{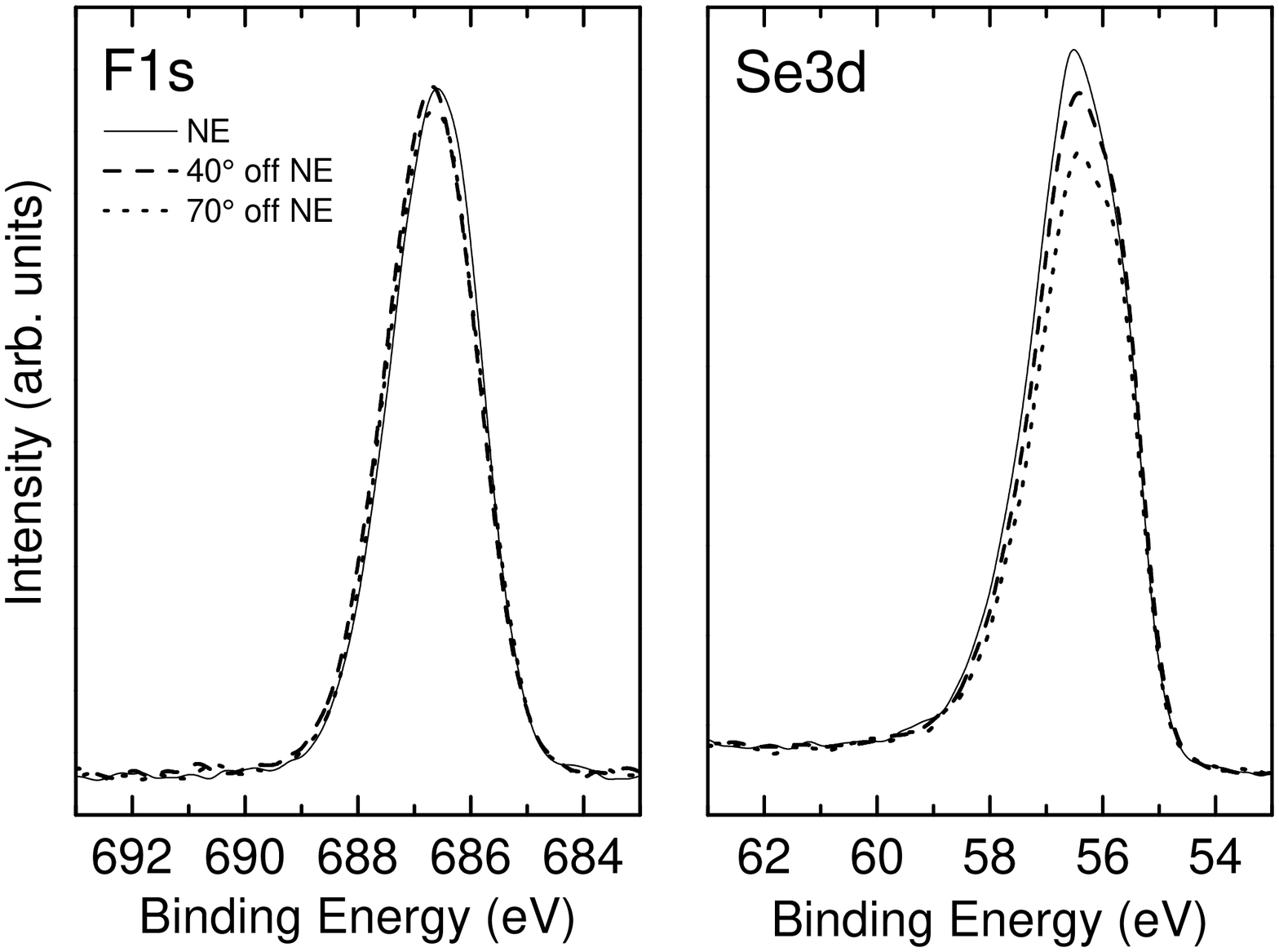}
\caption{\label{BS_F1s_Se3d}XPS spectra of the F~1$s$ and Se~3$d$ core
levels of {\BS} as a function of emission angle.}
\end{figure}

This conclusion is further corroborated, if one has a closer look
at the various core lines. The F~$1s$ line is expeditiously
treated (s. Fig.~\ref{BS_F1s_Se3d}). A single almost perfectly
symmetric line is observed at a binding energy of about 686.6\,eV.
There exist three crystallographically inequivalent lattice sites
for the fluorine atoms whose P--F bond length and angles, however,
do not much differ. In addition, the distance of the PF$_6$
complexes to the TMTSF stacks is very large. This excludes a
notable chemical shift of the binding energies. Since the PF$_6$
counter ions do not much hybridize with the TMTSF molecules and
thus do not participate in forming delocalized conduction bands
one would not expect any asymmetry of the F~1$s$ line as well.
Turning to the XPS spectrum of the Se~3$d$ doublet shown in
Fig.~\ref{BS_F1s_Se3d} we only see one single line at about
56.5\,eV binding energy because the spin-orbit splitting is too
small to be resolved. For similar arguments as above possible
chemical shifts in the binding energies of the four inequivalent Se
atoms should not be important. The bonding lengths and environment
of the Se atoms within the TMTSF molecule are quite the same and
their distances to the adjacent TMTSF molecules and PF$_6$ counter
ions are large. What is remarkable is the lack of a pronounced
asymmetric tail up to higher binding energies as it was observed
for {\TTF} and explained by the coupling of the photohole to the
conduction electrons. The Se atoms are located on the 1D conducting
stacks and a coupling of similar size as in {\TTF} would be
expected. This again manifests what we concluded already above from
the chemical analysis that apparently the {\BS} surfaces not at all
reflect bulk properties.

\begin{figure}
\includegraphics[angle=-90,width=8cm]{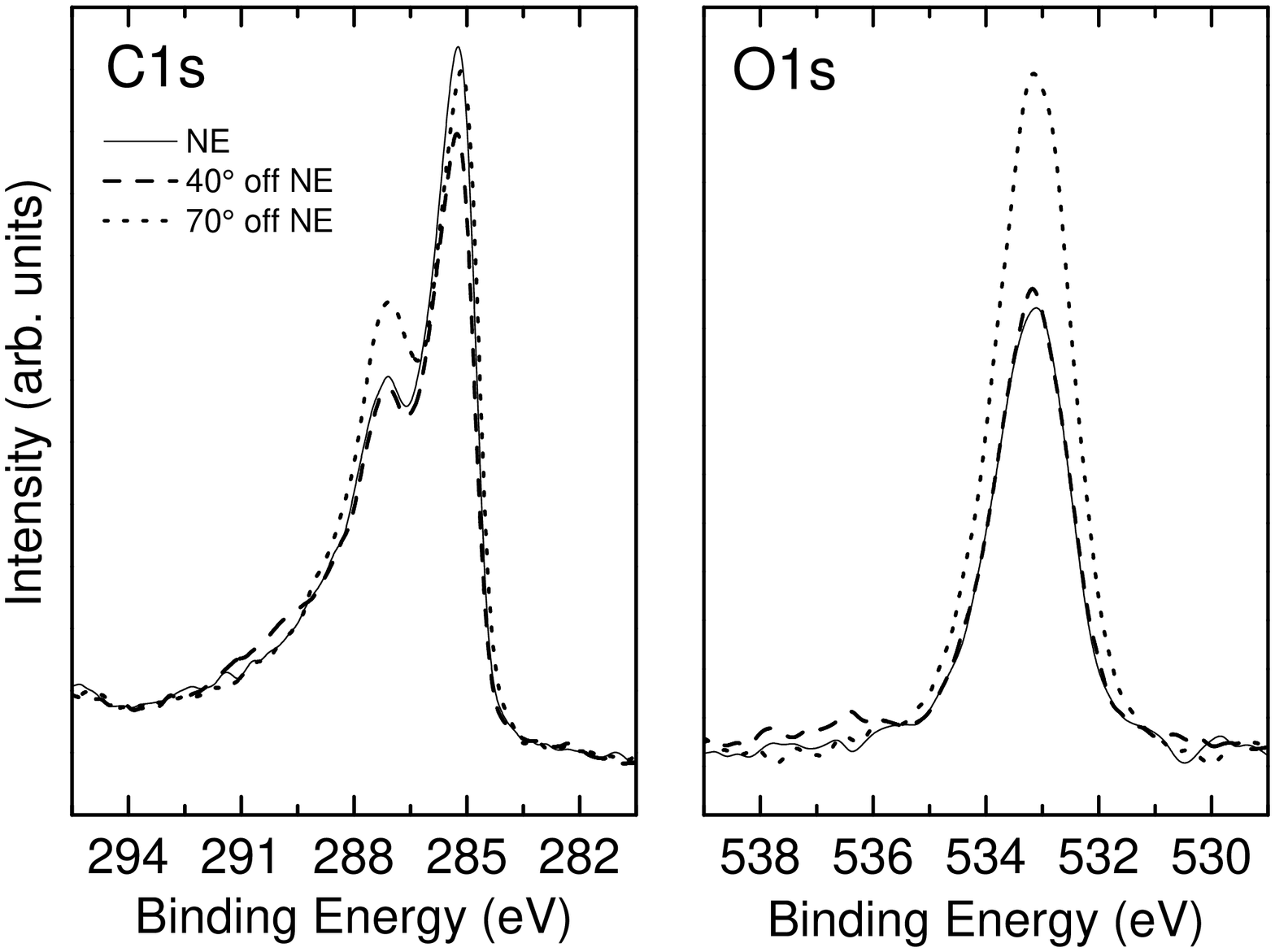}
\caption{\label{BS_C1s_O1s}XPS spectra of the C~1$s$ and O~1$s$ core
levels of {\BS} as a function of emission angle.}
\end{figure}
We refrain from a thorough discussion of the C~1$s$ line as it
overlaps with spectral weight due to Se~Auger electrons and
discuss it here only in the context of the angle dependence of the
various XPS lines (Figs.~\ref{BS_F1s_Se3d} and \ref{BS_C1s_O1s}).
In contrast to the F~1$s$ and Se~3$d$ lines the C~1$s$ line shows
a significant dependence upon variation of the emission angle. The
line is split into two components at about 284.7\,eV and 286.6\,eV
binding energy. The intensity of the latter increases notably at
the biggest off-normal emission angle of 70$^{\circ}$ thus
indicating a surface species. A similar even more pronounced
behavior is observed for the O~1$s$ line at about 533.2\,eV. We
draw two conclusions from those observations: Firstly, the
cleavage surface even if not as good as in the case of {\TTF} is
sufficiently well-defined to show angle dependencies at all. An
irregularly rough surface as generated by fracture (as opposed to
cleavage) of crystals would not display angular dependencies due
to the averaging of exit angles and shadowing
effects.\cite{Briggs90} Secondly, only part of the O signal can be
attributed to an O contamination on top of the topmost surface
layer. In the same way one can argue that also only part of the
C~1$s$ intensity is intrinsic due to the C atoms in the TMTSF
molecules, part stems from contamination of the topmost surface
layer, and part originates from C contamination built in the
crystal e.g. at microcracks.

\subsection{Crystalline surface order and ARPES}
\begin{figure}
\includegraphics[angle=-90,width=8cm]{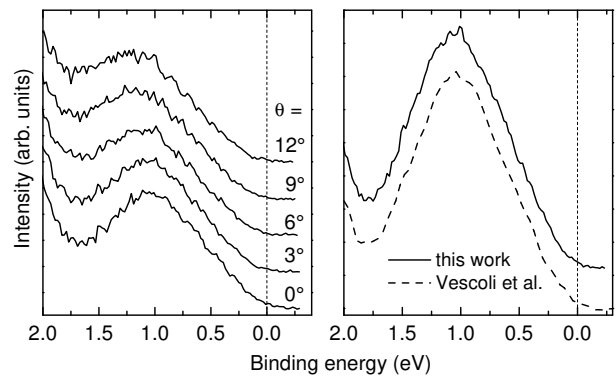}
\caption{\label{BS_UPS}Angle resolved (left panel) and angle
integrated (right panel) PES-spectra of {\BS}. The angle resolved
data were taken along the 1D axis.}
\end{figure}

In the light of the results of the preceeding paragraph it might
appear questionable whether one should anticipate long range
surface order for {\BS} at all. In any case, our attempts to see a
LEED pattern failed. Obviously, if there was any long range surface
order before, it is destroyed by the electron beam as in the case
of {\TTF}. Again we could use ARPES to reveal long range order by
the observation of dispersing electron states. ARPES spectra along
the 1D direction of {\BS} are shown in the left hand panel of
Fig.~\ref{BS_UPS}. In the energy range reaching to 2\,eV below
$E_F$ only one broad structure is observed with a maximum at about
1\,eV. This structure sits on a relatively high inelastic
background which artificially introduces a small shift to higher
binding energies. If one corrects the data for those secondary
electrons essentially no dispersion is seen. In the right hand
panel of Fig.~\ref{BS_UPS} we have summed up the ARPES spectra to
simulate an angle integrated spectrum which can be compared to
data previously published by Vescoli {\it et al.}\cite{Vescoli00}
The agreement is almost perfect. In the context of bulk-probing
optical and transport data in that paper the strong suppression of
spectral weight at $E_F$ as well as the specific power-law decay
of the leading edge towards the Fermi level was consistently
interpreted as evidence for a LL scenario. Only the exponent
governing the power-law decay would come out too high. However, it
was argued that this observation together with the absence of any
dispersion in related Bechgaard salts should rather be taken as
indirect manifestation of the LL phenomenology: while the bulk
properties can be reconciled within a standard LL picture,
impurities at the surface induce localization of the spin and
charge excitations which have to be described by a so-called
bounded LL.\cite{Voit00,Meden00} The finite length chains and the
thereby imposed boundary conditions would renormalize and thus
explain the unusual high power-law exponent. In the light of our
surface analysis we must however conclude that the measured
(AR)PES spectra do not represent intrinsic surface let alone bulk
properties of the Bechgaard salts.

Our reasoning on the Bechgaard salts in the context of ARPES
measurements may be parallelled and further corroborated by the
results published so far for the two-dimensional organic BEDT-TTF
salts. As in the Bechgaard salts their surfaces comprise either
anion or cation layers. Reconstruction/relaxation induced
structural modulations have been revealed on the surfaces of
various BEDT-TTF based compounds by STM.\cite{Ishida01} And again
PES fails to see a clear Fermi cut-off in the metallic BEDT-TTF
materials and notable dispersion of the electronic excitations
close at
$E_F$.\cite{Liu95a,Liu95b,Soederholm95,Soederholm97,Sekiyama97}
Thus one is led to speculate that it is indeed the influence of
surface effects, in particular their polar character, which in
many organic charge-transfer salts hampers the observation of the
electronic structure intrinsic for the bulk or a well-defined and
reproducible surface by means of PES. So PES often may only
pretend unconventional electronic behavior as it has been reported
previously.

\section{Conclusion}
In this paper we comprehensively studied the surfaces of two
organic charge-transfer salts, {\TTF} and {\BS}, in comparison.
Strong limitations regarding employable probing techniques are
imposed by their high sensitivity to chemical decomposition due to
electron and photon irradiation. We showed that against this
background x-ray induced photo\-emission spectroscopy is a
valuable diagnostic tool which does not destroy the surfaces
within reasonable time scales and provides information on surface
contaminations, surface stoichiometry, and even metallicity of the
surface. Thus it is possible to decide if such a surface most
probably exhibits intrinsic -- as is the case for {\TTF} -- or
extrinsic -- as is the case for {\BS} -- surface properties. In
how far intrinsic surfaces represent bulk properties, however, is
another question as we demonstrated for {\TTF}. There
photo\-emission spectra of the valence band showed clear
indication for renormalized electronic properties at the surface
with respect to the bulk. From our investigations we are able to
confirm the observation of generic one-dimensional features in
terms of spin-charge separation for {\TTF} while we can definitely
rule out unambiguous indications of Luttinger liquid behavior in
{\BS} as stated previously.

\begin{acknowledgments}
We gratefully acknowledge financial support by the Deutsche
Forschungsgmeinschaft (DFG) under CL 124/3-2.
\end{acknowledgments}


\begin{thebibliography}{3}
\bibitem{ann1}To determine the correct amount of carbon 10\,\% of the
area under the C~1$s$ peak was estimated to stem from underlying
Se~Auger excitations.
\end{thebibliography}
\end{document}